# Understanding Modality Preferences in Search Clarification

**Leila Tavakoli**
RMIT University
Australia
leila.tavakoli@student.rmit.edu.au

**Giovanni Castiglia**
Polytechnic University of Bari
Italy
g.castiglia@studenti.poliba.it

**Federica Calò**
Polytechnic University of Bari
Italy
f.calo8@studenti.poliba.it

**Yashar Deldjoo**
Polytechnic University of Bari
Italy
yashar.deldjoo@poliba.it

**Hamed Zamani**
University of Massachusetts Amherst
United States
zamani@cs.umass.edu

**Johanne R. Trippas**
RMIT University
Melbourne, Australia
j.trippas@rmit.edu.au

## ABSTRACT

This study is the first attempt to explore the impact of clarification question modality on user preference in search engines. We introduce the *multi-modal* search clarification dataset, *MIMICS-MM*, containing clarification questions with associated expert-collected and model-generated images. We analyse user preferences over different clarification modes of text, image, and combination of both through crowdsourcing by taking into account image and text quality, clarity, and relevance. Our findings demonstrate that users generally prefer *multi-modal* clarification over *uni-modal* approaches. We explore the use of automated image generation techniques and compare the quality, relevance, and user preference of model-generated images with human-collected ones. The study reveals that text-to-image generation models, such as *Stable Diffusion*, can effectively generate *multi-modal* clarification questions. By investigating *multi-modal* clarification, this research establishes a foundation for future advancements in search systems.

## 1 INTRODUCTION & CONEXT

**Context.** Effective communication between users and intelligent systems is of paramount importance in shaping enhanced search experiences. One common obstacle encountered by information retrieval (IR) systems is the inherent ambiguity present in human language. Clarification questions have emerged as an essential tool in search interactions, enabling users to refine their queries for more precise results [20]. However, their traditional *text-only* format often leaves users disengaged [18, 19].

While **multi-modal interactions** in IR systems are gaining traction, their impact, particularly through the *lens of clarifications*, remains under-explored. There is a significant gap in understanding how different modalities can enhance user experience. This paper seeks to bridge this gap. Consider a scenario where a user encounters the "Spectre" term and its symbol and seeks more information by querying it on an IR system; a text-only clarification might ask, "Did you mean Spectre comics or Spectre organisation?" Since the user lacks additional context about the symbol, selecting either option would be arbitrary. In contrast, a *multi-modal* clarification might display images of both options alongside the textual choices. Such a multi-modal approach provides an immediate visual context, making it easier for users to discern their initial intent. It also fosters a more interactive and intuitive user experience, reducing cognitive load and facilitating more informed decision-making.

The main question this paper seeks to answer is: "*can integrating visual presentations with text-only clarification panes[1] enhance the user experience, and if so, how significantly?*" To achieve this, we will employ a mixed-methods approach, combining qualitative interviews and quantitative measurements to gather both subjective user feedback and objective performance metrics. Through this approach, we aim to obtain a holistic understanding of user preferences, taking into account factors such as ease of use, efficiency, and overall satisfaction.

The main contributions of this paper are:

- **Introduction** of a novel *multi-modal* search clarification dataset, *MIMICS-MM*.
- **Conduct** an in-depth user study on preferences over different clarification modalities.
- **Analysis** of how visual and textual properties influence user preference.
- **Exploration** of the capabilities of text-to-image generation models in *multi-modal* search clarification.

**Related Work.** There has been a growing surge of interest in search clarification (i.g., Aliannejadi et al. [1], Braslavski et al. [2], Kim et al. [5], Tavakoli et al. [14]), however, the trajectory of most prior works, such as Aliannejadi et al. [1], Rao and Daumé III [8], Sekulić et al. [10], Zamani et al. [19] has been to enhance clarification modals' *effectiveness* in search systems. However, a major void remains concerning the **user perceptions** of different modalities in search clarifications.

While numerous studies have explored various facets of multi-modal systems, their focus has remained distinct from the area of our research. For instance, Yang et al. [17] introduced an online video recommendation system incorporating multi-modal fusion and relevance feedback. Zha et al. [21] proposed a Visual Query Suggestion system tailored for image search. Srinivasan and Setlur [11] explored utterance recommendations for visual analysis. In the field of robotics, Pantazopoulos et al. [6] integrated computer vision and conversational systems for socially enhanced robots, and Ferreira et al. [3] presented TWIZ, a multi-modal conversational task wizard. Remarkably, none of these studies explored *multi-modal* clarification questions in the context of search systems. This underscores a prominent research gap, emphasising the imperative for further investigation and innovation.

---

[1] According to Zamani et al. [18], a clarification pane comprises a multi-choice clarification question and a list of candidate answers. Hence, a *multi-modal* clarification pane contains both visual and textual content for each candidate answer.

**Research Approach.** In this research, we embark on the initial phase of comprehending the nuances of user preference in clarification modalities. Leveraging a crowdsourced user study, we investigate preferences over *textual*, *visual*, and *multi-modal* clarification panes. By systematically analysing user feedback, we can gain valuable insights into the advantages and limitations of each modality and the influential parameters. Sampling 100 query-clarification pairs from the *MIMICS* dataset introduced by Zamani et al. [18], we design both *visual* and *multi-modal* clarification panes, underpinned by expert annotations. To gauge the proclivity of users towards *multi-modal* panes over *uni-modal* ones, a post-task questionnaire is used, which also delves into the impact of image quality, text and image clarity and relevance.

A salient facet of our approach is to discern the feasibility of **automatically-generating images** for *text-only* clarification panes using models like Stable Diffusion [10]. We diligently assess the quality and relevance of such auto-generated images compared to human-collected ones.

**Key Findings.** Our research insights include:

- In the majority of cases (70-80%), users prefer *multi-modal* clarification panes over *visual-only* and *text-only* clarification panes. They also prefer *visual-only* clarification over *text-only* clarification in 54% of cases.
- Crowdworkers prefer *multi-modal* clarification panes as they are easier to understand, which helps users make better and faster decisions.
- Image quality, clarity, and relevance, in addition to text clarity, directly impact self-reported user perceptions.
- Text-to-image generation models, such as *Stable Diffusion* [9], are capable of automating image generation for creating *multi-modal* clarification panes.

In summary, this paper aims to contribute to the existing body of research on multi-modal search systems by investigating user preferences over different modalities for formulating clarification questions. We release the collected *multi-modal* search clarification dataset[2] and believe that this resource, in addition to the analyses conducted in this paper, will have significant implications for the design and development of more user-centric multi-modal search systems.

## 2 EXPERIMENTAL DESIGN

***Query and clarification panes sampling.*** We used the *MIMICS-Manual* dataset[3] to select textual clarification panes. We selected 100 queries randomly and their corresponding multi-choice clarification panes to create the *MIMICS-MM* dataset. The number of candidate answers in the clarification pane varies between two and five.

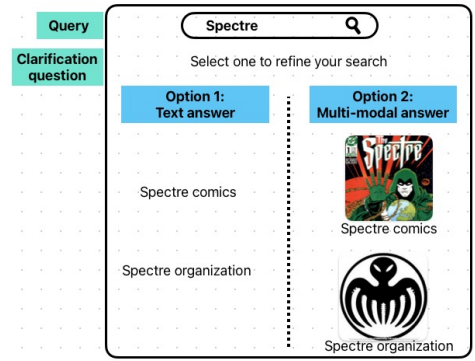

**Figure 1: An example of Task II (T vs. MM).**

***Clarification image collection.*** An expert annotator searched Google Images, assigning an image to each clarification pane's candidate answer. In total, 314 images were matched against 314 textual candidate answers.

***Experimental design.*** We conducted online experiments[4] on Amazon Mechanical Turk (AMT) collecting user preference labels through Human Intelligence Tasks (HITs).[5] We designed three tasks gathering judgements from AMT workers on user preferences over different modalities in search clarification. Detailed instructions with examples were prepared in plain English for each task to ensure the collected data was trustworthy. [6]

We ran pairwise comparisons as follows:

- Task I: *text-only* (T) vs. *visual-only* (V)
- Task II: *text-only* (T) vs. *multi-modal* (MM)
- Task III: *visual-only* (V) vs. *multi-modal* (MM)

Each HIT was assigned to at least three different AMT workers. For each labelling task, we used majority voting to aggregate the annotation. In case of disagreements, the HIT was opened again to more workers until a final majority vote label could be assigned. Figure 1 shows an example of Task II. A query and two modalities are displayed in each task through a Latin square design.

***Post-task questionnaire.*** We presented users with a post-task questionnaire assessing their presentation style preference and feedback, see Table 1. Thus, after inspecting the query, clarification pane, and candidate answers, workers indicated which presentation they preferred (*Q1*). The second question (*Q2*) presented checkboxes with options about the text and images' clarity, quality, and relevance. Workers were asked three more questions to obtain their motivation behind which modality was easier to understand (*Q3*), which helped them make better decisions (*Q4*) and faster decisions (*Q5*) on a 5-point scale. For example, in Task II, labels 1-2 means *text-only* modality is preferred, label 0 means they have no preference, and labels 4-5 means *multi-modality* is preferred.

***Quality assurance.*** Each task contained a gold question (i.e., a question with a known answer). Workers who failed to answer the gold question were prohibited from completing other tasks, and

---

**Table 1: Post-task questionnaire.**

| Post-task questionnaire |
| --- |
| 1) Which presentation do you prefer? |
| 2) Why did you pick the previous option? |
| 3) Which presentation did you find easier to understand? |
| 4) Which presentation helped you make a good decision? |
| 5) Which presentation helped you make the decision faster? |

**Table 2: Pairwise preference for clarification modality (%)**

| Task | Prefer text (T) | Prefer visual (V) | Prefer multi-modal (MM) | No preference |
| --- | --- | --- | --- | --- |
| **T vs. V** | 39[†] | 54[†] | NA | 7[†] |
| **T vs. MM** | 17[†] | NA | 79[†] | 4[†] |
| **V vs. MM** | NA | 17 | 71[†] | 12 |

[†] Significantly different from the other two preferences (Tukey HSD test, $p<0.05$).

their answers were removed. We also manually checked 10% of submitted HITs per task. Invalid submissions were removed, and the workers were denied from completing subsequent tasks. We then opened those HITs to other workers. We completed AMT pilot tasks[7] to analyse the task's flow, acquire users' feedback, check the quality of collected data, and estimate the required time to finish each task and a fair pay rate.

*Workers.* Only workers based in Australia, Canada, Ireland, New Zealand, the United Kingdom, and the United States, with a minimum HIT approval rate of 98% and a minimum of 5,000 accepted HITs, were allowed to participate. This maximised the collected data quality and the likelihood that workers were either native English speakers or had a high level of English.[8] Each HIT was assigned to at least three different AMT workers, enabling us to use an agreement analysis measure.

*Image generation for* **multi-modal** *clarification.* We employed two text-to-image generation models, *Stable Diffusion* [9] and *Dall·E 2* [7], for generating images for *multi-modal* candidate answers. *Stable Diffusion*[9] is a neural text-to-image model that uses a diffusion model variant called the latent diffusion model. It is capable of generating photo-realistic images given text input. Dall·E 2, created by OpenAI, generates synthetic images corresponding to an input text. Our input to generate a corresponding image to a candidate answer of a clarification pane was the concatenation of the query and the candidate answer text. This input generated all corresponding images for all candidate answers (two images were generated by two employed models per candidate answer).

*Comparing human-collected versus model-generated images.* Here, we conducted a manual annotation. We investigated the relevance of the generated images to the tests to evaluate how well the model-generated images present the text compared to the human-collected ones. We also compared the images' quality and assessed user preference over generated and collected images. Three annotators with proficient English and a higher degree completed the labelling. Each annotator labelled 314 generated images. We collected all annotations, aggregated them, and if there was a disagreement, majority voting was used for the final label. We showed the query concatenation and the candidate answer in the text, and the corresponding generated image to the annotators (similar to *Q1*, Figure 1). We asked annotators if the image was relevant to the text on a binary scale (i.e., label 1 means relevant). This label was similar to the label collected for the human-collected images during the crowdsourcing (*Q2*, Table 1). Then, we showed the collected image for the same text from the crowdsourcing part and asked the annotators to compare the quality of generated and collected images regardless of the presented text on a 3-point scale.[10] Finally, the annotators had to indicate their preferred image between two images presented for the text shown on a 3-point scale.[11]

## 3 RESULTS

We examined how different clarification modalities, as well as properties of both images and text, influence user preference. Following this, we explored the feasibility of automating the visual clarification modality.

*User preference and clarification modality.* We first investigated user preferences over the clarification modality in each pairwise comparison (i.e., *text-only* vs. *visual-only*, *text-only* vs. *multi-modal*, and *visual-only* vs. *multi-modal*). We performed Tukey honestly significant difference (HSD) [16][12] to help us determine, for instance, if the number of users who preferred *multi-modal* over *text-only* was significantly higher or not. Table 2 indicates the percentage of user preference in each pairwise clarification modality comparison. In Task I, where the workers needed to indicate their preferences between the *text-only* and *visual-only* clarifications, we observed that 54% of the workers preferred *visual-only* over *text-only* clarification panes. In Tasks II and III, where the workers needed to choose between *uni-modal* and *multi-modal* clarification panes, the workers strongly preferred *multi-modal* clarification panes, no matter whether the *uni-modal* clarification pane was *text-only* or *visual-only*. The workers' preferences were significantly different from other options indicating that in 70-80% of the cases, a *multi-modal* clarification was preferred. This clearly indicates that regardless of text and image quality and clarity, presenting a *multi-modal* clarification question in a search scenario was preferred by the users.

*Post-task questionnaire analysis.* We asked the workers to explain if the text/image clarity and relevance, and image quality impacted their preferences. We calculated the Pearson correlations between the workers' preferences and the characteristics of the clarification modalities in each Task. We observed that the calculated correlations for image quality, image/text clarity and image/text relevance were significantly different from each other in each task according to Tukey HSD test, $p<0.05$. In Task I, we observed a positive correlation ($\rho$=0.476) between user preference (i.e., preferring *visual-only* clarifications over *text-only* ones) and image quality.

---

[7] Pilot study was conducted in February 2022.

[8] Each worker could perform 25 tasks (a portion used for each launch). They had a five-minute time limit to finish the task and were compensated with 0.74 USD per HIT.

[9] We used the Diffusers library available at https://github.com/huggingface/diffusers.

[10] The quality of the model-generated image is higher *(2)*, are the same *(1)*, or the human-collected image has a higher quality *(0)*.

[11] Preference labels were between 2–0 where preferred the model-generated image *(2)*, had no preference *(1)*, or preferred the human-collected image *(0)*.

[12] The Tukey HSD test is a post hoc test used when there are equal numbers of subjects in each group for which pairwise comparisons of the data are made [12].

**Table 3: Motivations behind user preference (%).**

| Motivation | T vs. V | | T vs. MM | | V vs. MM | |
| | Prefer T | Prefer V | Prefer T | Prefer MM | Prefer V | Prefer MM |
|---|---|---|---|---|---|---|
| Easier to understand | 25 | 31 | 7 | 61 | 6 | 67 |
| Better decision | 22 | 36 | 6 | 68 | 3 | 67 |
| Faster decision | 27 | 36 | 10 | 62 | 6 | 66 |
| None of above | 8 | 12 | 4 | 9 | 9 | 5 |

**Table 4: Comparison of human-collected and model-generated search clarification pane images.**

| Collection Method | Relevance | Image Quality[2] | Image Preference[3] |
|---|---|---|---|
| Human-Collected | 96% | 42.7% | 60.2% |
| Model-Generated[1] | 87% | 20.7% | 12.7% |

[1] *Stable Diffusion* Model.

[2] 36.6% of users indicated the quality of the generated and collected images were the same.

[3] 27.1% of users indicated no preference over the generated and collected images.

There was also a strong positive correlation ($\rho$=0.677) between user preference and image clarity, and user preference had a strong negative correlation ($\rho$=-0.686) with text clarity. The same correlation trends and orders were observed for the user preference (i.e., preferring *multi-modal* clarifications over *text-only* ones) with image quality ($\rho$=0.458), image clarity ($\rho$=0.626) and text clarity ($\rho$=-0.627) in Task II. However, in Task III, the user preference (i.e., preferring *multi-modal* clarifications over *visual-only*) had correlations only with the text clarity ($\rho$=0.505) and image clarity ($\rho$=-0.301). The collected feedback from workers shows that the text and the image in more than 95% of clarification panes were relevant, explaining low correlations between user preference and the relevance of the text and the image.

In the pairwise comparison between *multi-modal* and *visual-only* clarification panes, although the collected images for the clarification panes were relevant for nearly all cases, the workers preferred *multi-modal* clarification panes over the *visual-only* ones when the images were not clear. The text helped them understand the candidate answers to the clarification panes. The users preferred *visual-only* clarifications in more than 54% of cases when the text clarity was low, and the image quality and clarity were high. However, the text and image were relevant in most cases.

We also asked users whether the preferred modality was easier to understand and helped them make better and faster decisions. Table 3 shows the user preferences in each pairwise modality. We see when users preferred *visual-only* clarification panes over *text-only* ones, 31% of users believed that the *visual-only* clarification panes were easier to understand, and the *visual-only* modality helped 36% of users make better and faster decisions. When comparing *multi-modal* clarification panes with *text-only* and *visual-only* clarification panes, between 60 to 70% of users believed that *multi-modal* clarification panes were easier to understand and helped them make better and faster decisions. Table 3 shows that there were small groups of users whose motivations behind their preferences were not listed in our questions.

**User preference and impact of visual aspects.** We investigated the impact of visual aspects of the collected images on user preference over the clarification modality. We extracted the visual aspects of the images using *OpenIMAJ* [4], a tool for multimedia content analysis. The nine investigated visual aspects were *brightness*, *colourfulness*, *naturalness*, *contrast*, *RGB contrast*, *sharpness*, *sharpness variation*, *saturation*, and *saturation variation* [15]. We calculated the point-biserial correlation [13][13] between the visual aspects of images and user preferences, the image quality and the image clarity. The average value of each aspect was calculated across all candidate answers for each clarification pane. Therefore, one value was obtained per visual aspect for every clarification pane. There was a low correlation between the image's visual aspects and user preference, including the image quality and clarity that the workers judged.

To further explore the impact of visual aspects of images on user preference, we developed a feature-level attribution explanation to rate the image's visual characteristics based on their user preference. We utilised the Gini importance of the random forest with visual aspects as the input and target label user preference. [14] We performed this analysis for Tasks II, and the results indicate that *brightness*, *naturalness*, *RGB contrast*, *sharpness variation*, and *saturation variation*, among other studied aspects, accounted for more than 65% of the differences in user preferences. In particular, *brightness* and *naturalness* were the two most important visual features.

**Automatic image generation for clarification panes.** Finally, we investigated if generating the corresponding images to the candidate answers could be automated. First, we compared the visual aspects of the generated images with the collected ones. We observed that the generated images had relatively the same visual aspects as the collected ones. However, the *Stable Diffusion* model generated images with more similar sharpness to the human-collected images compared to *Dall·E 2*.

In the second step, we focused on the images generated by the *Stable Diffusion* model that generated images more similar to human-collected ones in terms of visual characteristics. We compared model-generated images with human-collected ones regarding image relevance, quality, and user preference (Table 4). Our analysis shows that 87% of generated images were relevant to the text (compared to 96% of human-collected images that were relevant). Only 20.7% of the generated images had a higher quality than human-collected ones. However, more than 57% (20.7% + 36.6%) of model-generated images had higher or equal qualities than collected ones. Although these observations indicate the acceptable performance of the *Stable Diffusion* model in generating relevant and high-quality images for clarification panes in search engines, only 12.7% of the generated images were preferred over the human-collected images. Table 4 also indicates that less than 40% (12.7% + 27.1%) of the users either preferred the generated images or had no preferences over the generated and collected images (same preference).

The annotators preferred the collected images over ~60% of model-generated images. This observation was expected as the collected images were gathered through online searching to select the most appropriate images, while a text-to-image model generated an image from only text. However, the *Stable Diffusion* model could generate relevant and high-quality images. As, in ~80% of cases, users preferred a *multi-modal* clarification pane over a *text-only* one; such a text-to-image model can ease and fasten the task of generating *multi-modal* clarification panes.

---

[13]The point-biserial correlation measures the relationship between a binary (i.e., user preference, image quality, and clarity) and a continuous variable (i.e., image aspects).

[14]Label 0 means T preferred over MM and label 1 means MM preferred over T.

## 4 CONCLUSIONS AND FUTURE WORK

We aimed to understand the impact of clarification question modality on user preference. We introduced a novel *multi-modal* clarification dataset, MIMCS-MM. We created three modalities of *text-only*, *visual-only*, and *multi-modal* (a combination of both) for clarification panes and presented them to users through crowdsourcing.

The research shows that users generally preferred *multi-modal* clarification panes over *text-only* and *visual-only* ones. Users found it easier to understand the information presented in *multi-modal* panes, which helped them make better and faster decisions. This implies that integrating text and visual elements improves comprehension and decision-making for users, particularly given that the models for generating clarifications are not yet performing optimally. The study identified that when images were clear and of high quality, users favoured *multi-modal* panes. Therefore, ensuring that the visual content provided in clarification panes is of good quality and easily understandable is crucial. We also showed that when the images were unclear and of low quality, users preferred *text-only* clarification panes, even if the images were relevant. This suggests that when visual content is inadequate, relying solely on text can be more effective in conveying the necessary information.

We also explored the task of automatically generating corresponding images for *text-only* clarifications to make them *multi-modal* clarifications. The results indicated that text-to-image generation models, such as *Stable Diffusion*, can produce high-quality and relevant visual content. This discovery indicates that automated generation techniques can produce *multi-modal* panes for search clarifications. Nonetheless, it is crucial to note that these methods have not yet achieved the ability to completely replicate human annotation when gathering relevant images for *text-only* clarification panes. Users still strongly prefer images collected by humans rather than those generated by models.

Our objective in this study was to gain insight into user preferences regarding different clarification modalities in a search scenario rather than examining the impact of clarification modality on search performance. As a result, we acknowledge that the participants in our study were not in a genuine search situation.

In our research, we recognise the potential impact of the dataset size. However, the statistically significant differences observed in our analysis form a reliable foundation for drawing valid conclusions. We have utilised robust statistical techniques to ensure the credibility of our findings, and it is unlikely that the observed effects are solely due to random chance.

The conducted study suggests several research paths for the future, including investigating the impact of clarification modality on search performance in real search situations, creating a more comprehensive dataset containing various aspects of queries to explore clarification modality further, developing advanced *multi-modal* language models to determine the most effective modality in different scenarios, investigating the impact of factors like user demographics, task complexity, and content characteristics, improving image generation techniques to produce more preferable images, and lastly, exploring alternative modalities beyond text and images, such as audio or interactive elements.

*on Multimedia.* 15–24.